\documentclass[manuscript,screen]{acmart}

\usepackage{algorithm}
\usepackage{algpseudocode}
\usepackage{cleveref}
\usepackage{mleftright}
\usepackage{pgfplots}
\usepackage{physics}
\usepackage{subcaption}
\usepackage{listings}

\lstdefinestyle{myCStyle}{
  language=C,
  basicstyle=\ttfamily\footnotesize,
  numbers=none,
  numberstyle=\tiny,
  stepnumber=1,
  numbersep=5pt,
  backgroundcolor=\color{gray!10},
  keywordstyle=\color{blue},
  commentstyle=\color{green!50!black},
  stringstyle=\color{red},
  showstringspaces=false,
  tabsize=2,
  breaklines=true,
  breakatwhitespace=false,
  frame=single,
  captionpos=b,
  morekeywords={sunindextype,
                sunrealtype,
                sunbooleantype,
                N_Vector,
                SUNMatrix,
                SUNLinearSolver,
                SUNNonlinearSolver}
}

\usepgfplotslibrary{groupplots}
\pgfplotsset{
    compat=1.18,
    table/col sep=comma,
    legend style={font=\small},
    x label style={font=\small},
    y label style={font=\small},
    tick label style={font=\small}
}

\newcommand{\code}[1]{\texttt{#1}}
\newcommand{\rev}[1]{#1}

\algrenewcommand\algorithmicrequire{\textbf{Input:}}
\algrenewcommand\algorithmicensure{\textbf{Output:}}

\AtBeginDocument{%
  }

\setcopyright{usgovmixed}
\copyrightyear{2026}
\acmYear{2026}
\acmDOI{XXXXXXX.XXXXXXX}
\acmISBN{978-1-4503-XXXX-X/2018/06}

\begin{document}

\title{New Time Integrators and Capabilities in SUNDIALS Versions 6.2.0-7.4.0}

\author{Steven B. Roberts}
\email{roberts115@llnl.gov}
\orcid{0000-0002-7225-2501}
\affiliation{%
    \institution{Lawrence Livermore National Laboratory}
    \city{Livermore}
    \state{California}
    \country{USA}}

\author{Mustafa A\u{g}g\"{u}l}
\email{maggul@smu.edu}
\orcid{0000-0003-4013-9907}
\affiliation{%
    \institution{Southern Methodist University}
    \department{Department of Mathematics}
    \city{Dallas}
    \state{Texas}
    \country{USA}}
    
\author{Daniel R. Reynolds}
\email{dreynolds@umbc.edu}
\orcid{0000-0002-0911-7841}
\affiliation{%
    \institution{Southern Methodist University}
    \department{Department of Mathematics}
    \city{Dallas}
    \state{Texas}
    \country{USA}}
\affiliation{%
    \institution{University of Maryland Baltimore County}
    \department{Department of Mathematics and Statistics}
    \city{Baltimore}
    \state{Maryland}
    \country{USA}}
    
\author{Cody J. Balos}
\email{balos1@llnl.gov}
\orcid{0000-0001-9138-0720}

\author{David J. Gardner}
\email{gardner48@llnl.gov}
\orcid{0000-0002-7993-8282}

\author{Carol S. Woodward}
\email{woodward6@llnl.gov}
\orcid{0000-0002-6502-8659}
\affiliation{%
    \institution{Lawrence Livermore National Laboratory}
    \city{Livermore}
    \state{California}
    \country{USA}}

\renewcommand{\shortauthors}{Roberts et al.}

\begin{abstract}
    SUNDIALS is a well-established numerical library that provides robust and efficient time integrators and nonlinear solvers.
    This paper overviews several significant improvements and new features added over the last three years to support scientific simulations run on high-performance computing systems.
    Notably, three new classes of one-step methods have been implemented: low storage Runge--Kutta, symplectic partitioned Runge--Kutta, and operator splitting.
    In addition, we describe new time step adaptivity support for multirate methods, adjoint sensitivity analysis capabilities for explicit Runge--Kutta methods, additional options for Anderson acceleration in nonlinear solvers, and improved error handling and logging.
\end{abstract}

\begin{CCSXML}
<ccs2012>
   <concept>
       <concept_id>10002950.10003705.10003707</concept_id>
       <concept_desc>Mathematics of computing~Solvers</concept_desc>
       <concept_significance>500</concept_significance>
       </concept>
   <concept>
       <concept_id>10010405.10010432.10010442</concept_id>
       <concept_desc>Applied computing~Mathematics and statistics</concept_desc>
       <concept_significance>500</concept_significance>
       </concept>
   <concept>
       <concept_id>10011007.10011074.10011075.10011077</concept_id>
       <concept_desc>Software and its engineering~Software design engineering</concept_desc>
       <concept_significance>300</concept_significance>
       </concept>
   <concept>
       <concept_id>10002950.10003714.10003727.10003728</concept_id>
       <concept_desc>Mathematics of computing~Ordinary differential equations</concept_desc>
       <concept_significance>500</concept_significance>
       </concept>
   <concept>
       <concept_id>10002950.10003714.10003727.10003730</concept_id>
       <concept_desc>Mathematics of computing~Differential algebraic equations</concept_desc>
       <concept_significance>300</concept_significance>
       </concept>
 </ccs2012>
\end{CCSXML}

\ccsdesc[500]{Mathematics of computing~Solvers}
\ccsdesc[500]{Applied computing~Mathematics and statistics}
\ccsdesc[300]{Software and its engineering~Software design engineering}
\ccsdesc[500]{Mathematics of computing~Ordinary differential equations}
\ccsdesc[300]{Mathematics of computing~Differential algebraic equations}

\keywords{Numerical software, time integration, high-performance computing, operator splitting, Runge--Kutta methods}

\received{}
\received[revised]{}
\received[accepted]{}

\maketitle

\section{Introduction}
% 1 page, Steven

SUNDIALS, the SUite of Nonlinear and DIfferential/ALgebraic equation Solvers, is a collection of six packages for numerically solving systems of ordinary differential equations (ODEs), differential-algebraic equations (DAEs), and nonlinear, algebraic equations \cite{hindmarsh2005sundials,gardner2022sundials}.
The CVODE and IDA packages provide implementations of linear multistep methods for ODEs and DAEs, respectively. The CVODES \cite{serban2005cvodes} and IDAS packages additionally supply forward and adjoint sensitivity analysis capabilities.
The ARKODE package \cite{reynolds2023arkode} provides implementations of  several classes of one-step integrators including explicit, implicit, and implicit-explicit Runge--Kutta methods as well as multirate methods.
Finally, the KINSOL package contains solvers for nonlinear algebraic systems of equations.

In this paper, we present a significant expansion of the features and capabilities of SUNDIALS introduced between version 6.2.0 (released August 2022) and 7.4.0 (released June 2025).
In the ARKODE package, where the majority of the developments occurred, three new classes of methods were added.
The first is low storage Runge--Kutta (LSRK) methods \cite{KetchesonHighlyEfficientSSP, CondeSSPEmbeddings, sommeijer_rkc_1998, meyerSecondorderAccurateSuper2012, meyerStabilizedRungeKutta2014a} for memory constrained applications that require methods with a small memory footprint to solve systems of initial value problems (IVPs).
% Low storage implementations of strong stability preserving (SSP) Runge--Kutta methods are also included.
The second is symplectic partitioned Runge--Kutta (SPRK) methods \cite{hairerWarner2006geometric,sanzserna2016symplectic,sofroniou2003increment} which enhance SUNDIALS' offering of conservative methods for the important class of Hamiltonian systems. To our knowledge, this is the first general-purpose C/C++ implementation of SPRK methods enabled for high-performance computing.
Finally, operator splitting methods \cite{mclachlan2002splitting,blanes2024splitting} are a key addition for coupling SUNDIALS (or other) integrators together for multiphysics or multimodel simulations.
For systems with dynamics evolving on disparate time scales, the multirate methods in ARKODE now offer full time step adaptivity at each time scale for automatic error control and improved computational efficiency.
ARKODE has also been extended to enable adjoint sensitivity analysis for explicit Runge--Kutta methods, to support the growing need for differentiable simulations and ODE constrained optimization.
In the KINSOL package, the implementation of Anderson acceleration now supports varying the damping and depth parameters during the iteration for more robust and efficient nonlinear solves \cite{chen2024non,evans2020proof,pollock2021anderson}.
Finally, there have been improvements to error handling and logging across all of SUNDIALS leading to greater uniformity between packages and easier analysis of integrator and solver performance.

% \textcolor{red}{Should make a case for why these features are useful.  Convince a referee that this paper deserves to be published.}

All of the new SUNDIALS features leverage the existing library infrastructure, including the abstract interfaces for vector, matrix, linear solver, and nonlinear solver objects. SUNDIALS provides several implementations of these objects to support computations using shared memory parallelism, distributed memory parallelism, and GPU acceleration. 
Alternatively, users may also provide their own implementations allowing SUNDIALS to operate on application-specific data structures or leverage specialized algebraic solvers.
This object-oriented design enables SUNDIALS to be applied to problems ranging from small-scale experiments on a laptop to complex scientific simulations run on large-scale high-performance computing (HPC) systems \cite{reynolds2023arkode,balos_sundials_2025}.

SUNDIALS is written in C, and, as of SUNDIALS v7.0.0, requires a subset of the C99 standard.
The library also includes interfaces for utilizing SUNDIALS from C++ and Fortran codes.
SUNDIALS is released under the BSD 3-clause license and can be freely downloaded from \url{https://github.com/LLNL/sundials}.
% Additional information can be found at \url{https://computing.llnl.gov/projects/sundials}.
The remainder of this paper is organized as follows.  In \cref{sec:features} we discuss the new features added to SUNDIALS, in \cref{sec:results} we present some numerical results that exhibit many of the new features, and in \cref{sec:conclusions} we furnish some concluding remarks on the work.

\section{New Features}
\label{sec:features}
This section gives high-level overviews of the new features added to SUNDIALS in versions 6.2.0-7.4.0.  
Greater detail is provided in the cited references and the relevant SUNDIALS User Guides \cite{arkodeDocumentation,kinsolDocumentation}.

\subsection{Low Storage Runge--Kutta (LSRK) Methods}

In general, Runge--Kutta methods are considered ``low storage'' if \rev{they require only relatively few internal registers for stage storage (generally significantly fewer than the number of stages)}.  These methods are often of interest in memory constrained applications or computing environments.
%While many LSRK methods have been introduced in recent decades \cite{Higueras2021SPPLSRK, Cavaglieri2015IMEXLSRK, Higueras2019sspLSRKexp}, in order to support GPU-based hyperbolic and parabolic problems, we implemented two such families in the new LSRKStep module in ARKODE (added in SUNDIALS v7.2.0). 
The new LSRKStep module in ARKODE (added in SUNDIALS v7.2.0) includes two families of low storage methods designed for initial-value problems of the form 
\begin{equation} \label{eq:ode}
    \dot{y}(t)=f(t,y(t)), \quad y(t_0)=y_0, \quad t \in [t_0, t_f],
\end{equation}
where $t$ is the independent variable, $y$ is the state vector of dependent variables, and $\dot{y}$ denotes $\dv{y}{t}$.

The first family consists of the explicit strong stability preserving (SSP) Runge–Kutta methods of orders two through four from \cite{KetchesonHighlyEfficientSSP}. SSP methods are widely used in applications that arise from hyperbolic conservation laws, such as compressible or incompressible flows, and this family provides methods with optimal SSP time step restrictions. Step size adaptivity with these methods is achieved through the embedding coefficients provided in \cite{CondeSSPEmbeddings}. The second-order method may be formulated using any number of stages $s\ge 2$, the third-order method may be formulated for any $s=k^2$ with $k \ge 2$, and the fourth-order method fixes the number of stages to 10. Regardless of the number of stages chosen, all these methods only require storing two internal stages at a time.

The second family of methods included in LSRKStep are the so-called Super Time Stepping (STS) methods of order two: Runge--Kutta--Chebyshev (RKC) \cite{sommeijer_rkc_1998} and Runge--Kutta--Legendre (RKL) \cite{meyerSecondorderAccurateSuper2012, meyerStabilizedRungeKutta2014a}. These explicit methods are designed for problems where the Jacobian of the right-hand side function in \cref{eq:ode}, $\pdv{f}{y}$, has eigenvalues along the negative real axis. This arises in heat diffusion, astrophysical magnetohydrodynamics, and other applications with parabolic operators. 
Both RKC and RKL evolve $y_{n-1}\approx y(t_{n-1})$ to $y_n\approx y(t_n)$ with a time step $h_n = t_n-t_{n-1}$ using an update of the form
\begin{equation}
  \label{eq:RKC2}
  \begin{split}
    z_0 &= y_{n-1},\\
    z_1 &= z_0 + h_n \tilde{\mu}_1 f(t_{n-1}, z_0),\\
    z_j &= \mu_j z_{j-1} + \nu_j z_{j-2} + (1-\mu_j-\nu_j)z_0 
    + h_n \tilde{\mu}_j f(t_{n-1,j-1}, z_{j-1}) + h_n \tilde{\gamma}_j f(t_{n-1}, z_0), \quad j=2,\ldots,s,\\
    y_{n} &= z_s,
  \end{split}
\end{equation}
where $z_j$ are the internal stages, and the coefficients $\mu_j$, $\tilde{\mu}_j$, $\nu_j$, and $\tilde{\gamma}_j$ are method- and stage-number-specific.

% As seen in \cref{eq:RKC2}, these methods allow an arbitrary number of stages, which are used to enhance the extent of their linear stability region along the negative real axis: $h_n \rho \approx 0.81 s^2$ for RKC and $h_n \rho \approx (s^2+s-2)/2$ for RKL, where $h_n \rho$ is the maximum extent of this stability region.
As seen in \cref{eq:RKC2}, these methods allow for an arbitrary number of stages, and the number of stages determines the maximum extent of the linear stability region along the negative real axis: \rev{$0.653s^2$} for RKC and $(s^2+s-2)/2$ for RKL.
Currently, LSRKStep requires a user-provided function to compute an estimate of the largest eigenvalue of the Jacobian, $\rho \approx \lambda_{\textrm{max}}(\pdv{f}{y})$; however, unless a step is rejected due to excessive temporal error, the most recently computed estimate is reused across multiple steps before being recomputed to reduce computational cost.
Based on the estimate $\rho$, LSRKStep automatically selects the number of stages to satisfy the linear stability condition with the current time step \cite{sommeijer_rkc_1998, meyerSecondorderAccurateSuper2012, meyerStabilizedRungeKutta2014a}.
User-callable functions allow for customization of the frequency at which $\rho$ is recomputed, the safety factor applied when computing the number of stages, and the maximum number of stages allowed.

Time adaptivity for RKC and RKL methods is enabled via the local temporal error estimate, $\text{Est}_n$,  
\cite{sommeijer_rkc_1998}
\begin{equation}
  \label{eq:RKC-error-estimate}
  \text{Est}_{n} = \frac{1}{15}\left[12\left(y_{n-1}-y_{n}\right) + 6h_n\left(f(t_{n-1},y_{n-1})+f(t_{n},y_{n})\right)\right].
\end{equation}
The time steps themselves are determined by the temporal adaptivity strategies implemented in SUNDIALS \cite[Section 2.2.11]{arkodeDocumentation}, unless such a step would exceed the maximum number of allowed stages, in which case the time step is reduced.

%In addition to the existing ARKODE statistics, this new module can report how many times the dominant eigenvalue estimation is called, as well as the maximum number of stages used.

\subsection{Symplectic Partitioned Runge--Kutta (SPRK) Methods}
The SPRKStep module in ARKODE (added in SUNDIALS v6.6.0) implements SPRK methods \cite{hairerWarner2006geometric,sanzserna2016symplectic,sofroniou2003increment}
% \textcolor{red}{Cite the methods}
and is designed for separable Hamiltonian systems, $H(t,p,q) = T(t,p) + V(t,q)$, posed in the component partitioned IVP form,
\begin{equation}\label{eq:sprk_ivp}
    % \dot{y}(t) = 
    \begin{bmatrix}
        \dot{p}(t) \\
        \dot{q}(t)
    \end{bmatrix} =
    \begin{bmatrix}
        f_1(t,q(t)) \\
        f_2(t,p(t))
    \end{bmatrix},\quad
    y(t_0) =
    \begin{bmatrix}
        p_0 \\
        q_0
    \end{bmatrix},\quad
    f_1(t,q) = -\pdv{V(t,q)}{q},\quad
    f_2(t,p) = \pdv{T(t,p)}{p},\quad t \in [t_0, t_f].
\end{equation}
% -\frac{\partial V(t,q)}{\partial q} \\
% \frac{\partial T(t,p)}{\partial p} \\
When $H$ is a conserved quantity and a constant time step is used, SPRK methods approximately (with high accuracy) conserve $H$ for exponentially long times \cite{hairerWarner2006geometric}.

SPRKStep provides methods up to order 10 and two different algorithmic procedure options. 
% \textcolor{red}{Maybe state that both are explicit and whether they can use adaptive time steps.}
Both procedures are explicit and use fixed time steps.  
\Cref{alg:sprk_standard} requires 5 fewer floating point operations per time step and requires one fewer state vector to be stored in memory than \cref{alg:sprk_increment}. However, \cref{alg:sprk_increment} offers smaller roundoff accumulation (useful for long integration times) through compensated summation \cite{sofroniou2003increment}.
% \todo{How much less storage, multiple vectors or just a few scalars?} 
For either formulation, the method coefficients, $a$ and $\hat{a}$, are stored compactly in two separate arrays while the $c$ and $\hat{c}$ coefficients are computed dynamically.
Lagrange interpolation is used for dense output by default based on our own empirical data pointing to better conservation accuracy compared to Hermite interpolation.
However, users can switch to Hermite interpolation if desired.

% \textcolor{red}{Cite alg 1 the way you cited alg 2}

\begin{minipage}[t]{0.35\linewidth}
\begin{algorithm}[H]
    \centering
    \caption{SPRKStep Standard \cite{sofroniou2003increment}}\label{alg:sprk_standard}
    \begin{algorithmic}[1]
        \State Set $P_0 = p_{n-1},~ Q_1 = q_{n-1}$
        \For{$i = 1, \ldots, s$}
           \State $P_i = P_{i-1} + h_n \hat{a}_i f_1(t_{n-1} + \hat{c}_i h_n, Q_i)$
           \State $Q_{i+1} = Q_i + h_n a_i f_2(t_{n-1} + c_i h_n, P_i)$
        \EndFor
        \State Set $p_n = P_s, q_n = Q_{s+1}$
    \end{algorithmic}
\end{algorithm}
\end{minipage}
\hspace{0.05\linewidth}
\begin{minipage}[t]{0.53\linewidth}
\begin{algorithm}[H]
    \centering
    \caption{SPRKStep Increment \cite{sofroniou2003increment}}\label{alg:sprk_increment}
    \begin{algorithmic}[1]
        \State Set $\Delta P_0 = 0,~ \Delta Q_1 = 0$
        \For{$i = 1, \ldots, s$}
           \State $\Delta P_i = \Delta P_{i-1} + h_n \hat{a}_i f_1(t_{n-1} + \hat{c}_i h_n, q_{n-1} + \Delta Q_i)$
           \State $\Delta Q_{i+1} = \Delta Q_i + h_n a_i f_2(t_{n-1} + c_i h_n, p_{n-1} + \Delta P_i)$
        \EndFor
        \State Set $\Delta p_n = \Delta P_s, \Delta q_n = \Delta Q_{s+1}$
        \State Set $p_n = p_{n-1} + \Delta p_n, q_n = q_{n-1} + \Delta q_n$ \Comment{Kahan sums}
    \end{algorithmic}
\end{algorithm}
\end{minipage}
 
\subsection{Operator Splitting Methods}
\label{sec:operator_splitting}

In this subsection we discuss the two types of operator splitting methods added to ARKODE in SUNDIALS v7.2.0.

\subsubsection{Standard Operator Splitting}

The SplittingStep module in ARKODE is designed for IVPs of the form
\begin{equation} \label{eq:partitioned_OD}
    \dot{y}(t) = f_1(t,y(t)) + f_2(t,y(t)) + \dots + f_P(t,y(t)), \quad y(t_0) = y_0, \quad t \in [t_0, t_f],
\end{equation}
\rev{with the right-hand side additively split into $P > 1$ operators.}
Operator splitting methods \cite{mclachlan2002splitting,blanes2024splitting}, such as those implemented in SplittingStep, allow each \rev{operator} to be integrated separately, possibly with different numerical integrators or exact solution procedures.
% Coupling is only performed through initial conditions that are passed from the evolution of one operator to the next.
Despite the popularity of operator splitting in multiphysics simulations, there are few general-purpose, high-order software implementations available, e.g., \cite{guenter2024pythos,moayeri2025tost}.

\begin{algorithm}
    \caption{Operator splitting}\label{alg:splitting}
    \begin{algorithmic}[1]
        \For{$i = 1, \dots, r$} \Comment{Loop over each sequential method}
            \State{Set $y_{n,i} = y_{n-1}$} \Comment{Start the sequential method with the initial condition}
            \For{$j = 1, \dots, s$} \Comment{Loop over each stage}
                \For{$k = 1, \dots, P$} \Comment{Loop over each \rev{operator}}
                    \State{Let $t_{\text{start}} = t_{n-1} + \beta_{i,j,k} h_n$ and $t_{\text{end}} = t_{n-1} + \beta_{i,j+1,k} h_n$}
                    \State{Let $v(t_{\text{start}}) = y_{n,i}$}
                    \State{For $t \in [t_{\text{start}}, t_{\text{end}}]$ solve $\dot{v}(t) = f_k(t, v(t))$} \Comment{Evolve \rev{operator} $k$ from stage $j$ to $j+1$}
                    \State{Set $y_{n,i} = v(t_{\text{end}})$}
                \EndFor
            \EndFor
        \EndFor
        \State{Set $y_n = \sum_{i=1}^{r} \alpha_i y_{n, i}$} \Comment{The final solution is a linear combination of sequential method solutions}
    \end{algorithmic}
\end{algorithm}

To perform a step of size $h_n$, SplittingStep applies \cref{alg:splitting} where $s$ denotes the number of stages, while $r$ is the number of \textit{sequential methods}.
A sequential method starts from $y_{n-1}$ and evolves it through a sequence of subintegrations where the output of one subintegration is the input to the next.
All $r$ sequential methods are independent, and their final states are combined with a weighted sum to produce the overall operator splitting step $y_n$. The coefficients $\alpha \in \mathbb{R}^r$ define these weights and $\beta \in \mathbb{R}^{r \times (s + 1) \times P}$ determines the subintegration time intervals.
\rev{Users can provide their own coefficients or use built-in ones for many standard operator splitting methods, e.g., Lie--Trotter, parallel, Strang--Marchuk \cite{strang1968construction,Marchuk1968}, and Yoshida \cite{yoshida1990construction}.
We note Strang--Marchuk and Yoshida splittings are special cases of the family of triple jump methods \cite{CrGo:89} \cite[page 44]{hairerWarner2006geometric} which can attain arbitrarily high (even) order. Both triple jump and the closely related quintuple jump \cite{Suzuki:93} \cite[page 45]{hairerWarner2006geometric} methods are implemented in SplittingStep.}

An alternative and more mathematically conventional representation of the SplittingStep solution is
\begin{equation} \label{eq:operator_splitting}
    y_n = \sum_{i=1}^P \alpha_i \left(
    \phi^P_{\gamma_{i,s,P} h_n} \circ
    \phi^{P-1}_{\gamma_{i,s,P-1} h_n} \circ \dots \circ
    \phi^{1}_{\gamma_{i,s,1} h_n} \circ
    \phi^P_{\gamma_{i,s-1,P} h_n} \circ \dots \circ
    \phi^1_{\gamma_{i,s-1,1} h_n} \circ \dots \circ
    \phi^P_{\gamma_{i,1,P} h_n} \circ \dots \circ
    \phi^1_{\gamma_{i,1,1} h_n}
    \right)(y_{n-1})
\end{equation}
where $\gamma_{i,j,k} = \beta_{i,j+1,k} - \beta_{i,j,k}$ is the scaling factor for the time step, $h_n$, and $\phi^k_{h_n}$ is the flow map for \rev{operator} $k$:
\begin{equation} \label{eq:subintegration}
    \phi^k_{h_n}(y_{n-1}) = v(t_n),
    \text{ where }
    \dot{v}(t) = f_k(t, v(t))
    \text{ and }
    v(t_{n-1}) = y_{n-1}.
\end{equation}

Operator splitting allows for a great deal of flexibility in how the subintegrations in \cref{eq:subintegration} are computed.
To maintain this important property, SUNDIALS v7.2.0 introduced a \code{SUNStepper} interface to represent an arbitrary solution process for \cref{eq:subintegration}.
One \code{SUNStepper} is associated with each \rev{operator of \cref{eq:partitioned_OD}}. Utility functions are provided to convert any ARKODE integrator into a \code{SUNStepper}.
For example, one can use ARKODE's Runge--Kutta methods for the subintegrations, yielding a fractional step Runge--Kutta method \cite{spiteri2023fractional}.
Users can also supply their own \code{SUNStepper} implementations to wrap custom integrators or to analytically compute subintegration solutions.

\subsubsection{The Forcing Method}

The ForcingStep module in ARKODE is designed for IVPs of the form \cref{eq:partitioned_OD} with $P = 2$ operators.
A step from $t_{n-1}$ to $t_n$, with the forcing method implemented in ForcingStep is given in \cref{alg:forcing_method}. 
This approach resembles a Lie--Trotter splitting in that the \rev{operators} are evolved through a sequence of subintegrations from time $t_{n-1}$ to $t_n$.
However, the IVP for \rev{operator} two includes a ``forcing'' or ``tendency'' term $f_1^*$ to strengthen the coupling between \rev{operators}.
For this reason, it cannot be cast as a traditional operator splitting scheme as given in \cref{eq:operator_splitting}, and it is separate from the SplittingStep module.
\begin{algorithm}
    \caption{Forcing step}\label{alg:forcing_method}
    \begin{algorithmic}[1]
        \State{Let $v_1(t_{n-1}) = y_{n-1}$}
        \State{For $t \in [t_{n-1}, t_{n}]$ solve $\dot{v}_1(t) = f_1(t, v_1(t))$} \Comment{Evolve \rev{operator} 1 from $t_{n-1}$ to $t_n$}
        \State{Compute $f_1^* = \frac{v_1(t_n) - v_1(t_{n-1})}{h_n}$} \Comment{Compute the ``forcing'' term between \rev{operators}}
        \State{Let $v_2(t_{n-1}) = y_{n-1}$}
        \State{For $t \in [t_{n-1}, t_{n}]$ solve $\dot{v}_2(t) = f^*_1 + f_2(t, v_2(t))$} \Comment{Evolve \rev{operator} 2 with added forcing from $t_{n-1}$ to $t_n$}
        \State{Set $y_{n} = v_2(t_n)$}
    \end{algorithmic}
\end{algorithm}

ForcingStep is based on what is referred to as ``sequential-tendency splitting'' in \cite{donahue2018impact} and ``dribbling'' in \cite{wan2021quantifying}, and this method was found to be more efficient than other first order splittings in Earth system modeling 
due to the increased coupling.
The two subintegrations in \cref{alg:forcing_method} can be solved with an arbitrary integrator or exact solution procedure using the previously discussed \code{SUNStepper} interface.
Provided these are done to at least first order accuracy, the overall forcing method is also first order accurate.

% \begin{equation} \label{eq:forcing_method}
%     \begin{alignedat}{2}
%         v_1(t_{n-1}) &= y_{n-1},
%         & \qquad
%         v_2(t_{n-1}) &= y_{n-1}, \\
%         %
%         \dot{v}_1 &= f_1(t, v_1),
%         & \qquad
%         \dot{v}_2 &= f_1^* + f_2(t, v_2), \\
%         %
%         f_1^* &= \frac{v_1(t_n) - y_{n-1}}{h_n},
%         & \qquad
%         y_n &= v_2(t_n).
%     \end{alignedat}
% \end{equation}

\subsection{Multirate Adaptivity}

% 1 page, Dan \textcolor{blue}{(This still may be a bit too long -- suggestions on how to trim would be appreciated)}

We have upgraded (in SUNDIALS v7.2.0) the previously-released MRIStep module within ARKODE to support a broader range of multirate methods and enable temporal adaptivity, even at the slow time scale. MRIStep solves IVPs of the form
%MRIStep is the newest time-stepping module in ARKODE and targets multirate IVPs of the form
\begin{equation}
  \label{eq:MRIStep_IVP_3comp}
  \dot{y}(t) = f^{S}(t,y(t)) + f^F(t,y(t)), \quad y(t_0) = y_0, \quad t \in [t_0, t_f],
\end{equation}
where $f^S(t,y(t))$ corresponds with slow processes that should be integrated with a large time step, $H_n$, and where $f^F(t,y(t))$ corresponds with fast processes that should be integrated with smaller steps, $h_n \ll H_n$.  
To support adaptivity of both $H_n$ and $h_n$, an MRI method must include coefficients for both a time step solution, $y_n \approx y(t_n)$, and an embedding, $\tilde{y}_n\approx y_n$, that is computed using an identical structure as the final internal MRI stage.  Thus, as part of this upgrade, the suite of supported methods in MRIStep has been expanded to include embedded \rev{multirate infinitesimal GARK (MRI-GARK)} \cite{sandu_class_2019,roberts_fast_2022}, \rev{implicit-explicit multirate infinitesimal stage-restart (IMEX-MRI-SR)} \cite{fish_implicitexplicit_2024}, and \rev{multirate exponential Runge--Kutta (MERK)} \cite{luan_new_2020} methods, where the MERK embeddings were derived specifically for MRIStep.  \rev{The new methods, as well as extensive tests of the new multirate adaptivity features, are described in \cite{reynolds_efficient_2025}.}

%\textcolor{red}{CSW: I see a couple ways to shorten.  First, if decoupled and stepsize tolerance options are included in other documents, just cite them and remove the next two paragraphs (but keep the first sentence of the next paragraph).  If they are not, then I would suggest adding the content of the next two paragraphs to the ARKODE UG and cite it.  The level of detail of eth next two paragraphs far exceeds what we have had for the previous methods.}

MRIStep supports two types of multirate step size controllers: \emph{decoupled} and \emph{stepsize-tolerance} \cite{arkodeDocumentation}; both require an estimate of the slow temporal error, that we compute via the embedding as $\text{Est}_{n}^{S} = y_n - \tilde{y}_n$.
% As with other SUNDIALS integrators, this norm incorporates user-specified relative and absolute tolerances, and adaptivity is performed to ensure that this \emph{local error} satisfies $\varepsilon_n^S \le 1$.
The fundamental difference between the two families of multirate controllers is their level of coordination when evolving the fast and slow time scales. The \emph{decoupled} family combines separate single-rate adaptive controllers to individually select the fast and slow time steps, $h_n$ and $H_n$. The second family is designed for problems with more strongly coupled time scales such that fast error may accumulate to pollute the overall MRI step.  For these problems, the \emph{stepsize-tolerance} controllers still leverage individual single-rate controllers to adjust $h_n$ and $H_n$, but they also adaptively adjust the tolerance requested from the inner integrator when solving the fast subproblems.  We note that since both families focus on only one scale at a time, or the relationship between one time scale and the next-faster scale, they may easily be extended to problems with an arbitrary number of time scales, thereby supporting so-called ``telescopic'' MRI methods.

\subsection{Discrete Adjoint Sensitivity Analysis for Explicit Runge--Kutta Methods} 

% \mathbb{R} \times \mathbb{R}^N \times \mathbb{R}^{N_s} \to \mathbb{R}
For optimization problems we often have a functional, $g(t_f, y(t_f, p), p)$, for which we would like to compute the gradients $\dv{g}{y}$ and, optionally, $\dv{g}{p}$, where $p$ is a vector of system parameters, and $y(t, p)$ is the solution of the IVP,
\begin{equation}\label{eq:paramterized_ivp}
    \dot{y}(t, p) = f(t,y(t,p),p),\quad y(t_0, p) = y_0(p),\quad t \in [t_0, t_f].
\end{equation}
The CVODES \cite{serban2005cvodes} \rev{and IDAS packages in SUNDIALS have} long provided an approach for computing these gradients through \textit{continuous} adjoint sensitivity analysis (ASA).
% ASA in general is very efficient when $N_s$ is large and there are relatively few functionals because it only requires a single forward solution of \cref{eq:paramterized_ivp} and a backward integration of the adjoint system. 
Continuous ASA allows for solving \cref{eq:paramterized_ivp} with any numerical integration scheme and using time step and/or order adaptivity during both the forward and backward integration phases.
However, a drawback of continuous ASA is that it may not recover the exact gradients of the discrete-time problem that arise by discretizing \cref{eq:paramterized_ivp} with a numerical integration scheme \cite{sanzserna2016symplectic,tran2024properties,gholami2019anode,giles2000introduction}.
In the context of optimization problems, the inconsistent gradients can cause an optimizer to fail \cite{gholami2019anode,giles2000introduction}. 
As an alternative, we added a \textit{discrete} ASA capability in SUNDIALS v7.3.0 using any fixed-step explicit Runge--Kutta method supported by ARKODE. That is, given the $s$-stage explicit Runge--Kutta method,
\begin{equation}\label{eq:erk_scheme}
   z_i = y_{n-1} + h_n \sum_{j=1}^{i-1} a_{i,j} f(t_{n-1} + c_j h_n, z_j),
    \; i=1,\ldots,s,
    \qquad y_n = y_{n-1} + h_n \sum_{i=1}^{s} b_i f(t_{n-1} + c_j h_n, z_i),
\end{equation}
where $a_{i,j}$, $b_i$, and $c_i$ are the method defining coefficents, we implement the solution procedure,
\begin{subequations}
    \begin{align}
       \Lambda_i &= h_n \left(\pdv{f(t_{n,i}, z_i, p)}{y}\right)^* \left(b_i \lambda_{n+1} + \sum_{j=i+1}^s a_{j,i} \Lambda_j \right), \quad i = s, \dots, 1, \label{eq:erk_adjoint_Lambda} &
       \lambda_n &= \lambda_{n+1} + \sum_{j=1}^{s} \Lambda_j, \\
       \nu_i     &= h_n \left(\pdv{f(t_{n,i}, z_i, p)}{p}\right)^* \left(b_i \lambda_{n+1} + \sum_{j=i}^{s} a_{j,i} \Lambda_j \right), \label{eq:erk_adjoint_nu} &
       \mu_n     &= \mu_{n+1} + \sum_{j=1}^{s} \nu_j.
    \end{align}
\end{subequations}
% that follows from $\pdv*{g}{y_n} = \lambda_n$ and $\pdv*{g}{p} = \mu_n + \lambda_n^* \left(\pdv*{y_0}{p}\right)$ 
For a detailed derivation, refer to
\cite{sanduDiscrete2006,hager2000runge}.
The discrete adjoint provides the exact gradient of the discrete-time problem. Users of SUNDIALS must provide functions which perform the Jacobian-Hermitian-vector-products 
needed in \cref{eq:erk_adjoint_Lambda,eq:erk_adjoint_nu}; these can be analytic or computed through reverse-mode algorithmic differentiation tools. 
% In the future, we may extend the discrete ASA capability to allow time step adaptivity, relaxation \cite{bencomo2023discrete}, and other time integration methods available in SUNDIALS.

To support the discrete ASA capability, SUNDIALS v7.3.0 adds the \code{SUNAdjointStepper} interface to represent a generic adjoint solution procedure, which could be either discrete or continuous, and leverages the \code{SUNStepper} interface discussed in \cref{sec:operator_splitting}.
Storing and loading states from the forward IVP solution (the $z_i$ of \cref{eq:erk_scheme} in the case of the explicit Runge--Kutta methods) during the adjoint integration is handled through the new \code{SUNAdjointCheckpointScheme} interface. Currently SUNDIALS provides a ``fixed'' \code{SUNAdjointCheckpointScheme} implementation that stores states in-memory at a given fixed interval of time steps.
% Internally, \code{SUNAdjointCheckpointScheme} classes make use of a new (private) abstract hierarchical data storage class which handles saving the state data to a location defined by different class implementations. With SUNDIALS v7.3.0, only an in-memory storage implementation is provided.
% Future development may include more advanced checkpointing schemes as well as disk-based and multi-level storage of checkpoint states.

\subsection{Extensions to Anderson Acceleration}
% (1 page, David)
% v2.6.0 added Anderson acceleration 
% v2.6.2 added QR updating 
% v5.8.0 added damping and delay
% v6.1.0 added low sync
% v7.3.0 added variable damping and depth, reference Tango
% Damping \cite{chen2024non,evans2020proof} Chen "optimal", Evans heuristic
% Depth \cite{pollock2021anderson} heuristic depth based on the residual
% Filtering \cite{pollock2023filtering}

The KINSOL package in SUNDIALS provides a number of methods for solving systems of nonlinear algebraic equations. 
For problems of the from $G(u) = u$, the package includes fixed-point and Picard iterations with optional Anderson acceleration (AA) \cite{anderson1965iterative,walker2011anderson} to improve the rate of convergence.
%(\cref{alg:anderson}).
First introduced in SUNDIALS v2.6.0, these solvers have been steadily extended to include efficient QR updates, fixed damping and acceleration delay options, and low-synchronization orthogonalization methods \cite{lockhart2022performance}. 
See \cite[Section 7.2.12]{kinsolDocumentation} for more information on the AA implementation within KINSOL.

In SUNDIALS v7.3.0, we have added support for user control of the damping parameter and AA acceleration space size (depth) through the user-supplied callback functions in Listings \ref{lst:aa_damping} and \ref{lst:aa_depth}, respectively.
These new callback functions supply users with the necessary information to apply recently developed approaches for adaptively selecting the damping and depth parameters across solver iterations. For example, the methods presented in \cite{evans2020proof} and \cite{chen2024non} for selecting the damping strength based on the predicted benefit of acceleration (gain) or minimizing the nonlinear residual, respectively, showed notable improvements over using a fixed damping value. Methods for adapting the acceleration space size have also demonstrated better convergence compared to using a fixed depth. In \cite{pollock2021anderson} a heuristic approach based on the residual magnitude is employed while \cite{pollock2023filtering} presents a filtering method to control the condition number of the least-squares problem.

\begin{lstlisting}[style=myCStyle, caption={The required function signature for a user-supplied \code{KINDampingFn} function to set the AA damping parameter.}, label={lst:aa_damping}]
typedef int (*KINDampingFn)(long int iter,               // current iteration number
                            N_Vector u_val,              // current iterate
                            N_Vector g_val,              // fixed-point function evaluated at u_val
                            sunrealtype qt_fn[],         // values needed to compute the AA gain
                            long int depth,              // current depth
                            void *user_data,             // user-supplied pointer
                            sunrealtype *damping_factor) // output damping factor
\end{lstlisting}

\begin{lstlisting}[style=myCStyle, caption={The required function signature for a user-supplied \code{KINDepthFn} function to set the AA depth parameter.}, label={lst:aa_depth}]
typedef int (*KINDepthFn)(long int iter,                 // current iteration number
                          N_Vector u_val,                // current iterate
                          N_Vector g_val,                // fixed-point function evaluated at u_val
                          N_Vector f_val,                // current resiudal
                          N_Vector df[],                 // residual difference history
                          sunrealtype R_mat[],           // R matrix from the AA QR factorization
                          long int depth,                // current depth
                          void *user_data,               // user-supplied pointer
                          long int *new_depth,           // output depth
                          sunbooleantype remove_index[]) // which history vectors to remove
\end{lstlisting}

When applying KINSOL to a nonlinear diffusion problem relevant to fusion plasma applications \rev{in} \cite{gardner2024towards} \rev{and using the new features noted above}, variations of the approaches from \cite{evans2020proof} and \cite{pollock2021anderson} showed increased iteration robustness and reduced iteration counts compared to a relaxed fixed-point iteration and matched or bested the results from AA with fixed iteration delay, damping, and depth parameters selected using Bayesian optimization.

\subsection{Error Handling and Logging}

SUNDIALS v7.0.0 introduced a new suite-wide error handling interface along with improved error messages and a capability to log error, warning, informational, and debugging output to different file handlers. 
Previously, error handling was specific to each SUNDIALS package, and errors that occurred within shared SUNDIALS modules were not uniformly caught or logged.
Through new functions added to the \code{SUNContext} class, users can push custom error handler callback functions onto a stack, permitting layered error handling.
By default, SUNDIALS provides an error handler that simply logs errors, or an error handler that will terminate the program. 
Internally, many more error checks were added into the suite which has made errors easier to debug. 
These error checks rely on functions either returning a new \code{SUNErrCode} value, or by checking a ``last error'' value stored within the \code{SUNContext} (see Listing \ref{lst:sunchecks}). 
The return-code approach is used with all new functions added to SUNDIALS, while the ``last error'' approach is employed with existing functions that return a different value to maintain backwards compatibility.
The amount of error checking is configure-time selectable, allowing users to decide between more error checks or better performance.

\begin{lstlisting}[style=myCStyle, caption={Example of the new SUNDIALS error checking from a user's perspective.}, label={lst:sunchecks}]
// Every code that uses SUNDIALS must create a SUNContext.
SUNContext sunctx = SUNContext_Create(...);

// If a function does not return an error code, we check for errors using SUNContext_GetLastError
N_Vector v = N_VNew_Serial(2, sunctx); // Create a SUNDIALS serial vector of length 2.
SUNErrCode sunerr = SUNContext_GetLastError(sunctx);
if (sunerr) {  /* an error occurred, do something */ }

// If the function returns a SUNErrCode, we can check it directly
sunerr = N_VLinearCombination(...);
if (sunerr) {  /* an error occurred, do something */ }
\end{lstlisting}

All messages in SUNDIALS now flow through a new \code{SUNLogger} class. This class allows users to set file handlers for different classes of messages either through its API, or through environment variables at runtime.
Messages are output in a linear, structured, format (see Listing \ref{lst:logging_output}) that is both human- and machine-readable.
A Python-based utility to parse the output is distributed as part of SUNDIALS (see the \href{https://github.com/LLNL/sundials/tree/e941546af1a5b5e492dcac0a1872540e9961c556/tools/suntools}{\code{tools/suntools}} directory).
As a result, we have added many new logging statements to the suite, which can be enabled/disabled at configuration time. 
These additions enable SUNDIALS developers and advanced users to inspect the integrator and solver state at a granularity that was not previously possible, as demonstrated by the example scripts found within the \href{https://github.com/LLNL/sundials/tree/e941546af1a5b5e492dcac0a1872540e9961c556/tools}{\code{tools}} directory.

\begin{lstlisting}[style=myCStyle, caption={A snippet of logging output produced when using the ARKODE ARKStep module with the maximum logging level.}, label={lst:logging_output}]
[INFO][rank 0][ARKodeEvolve][begin-step-attempt] step = 1, tn = 0, h = 0.000102986025609508
[INFO][rank 0][arkStep_TakeStep_Z][begin-stage] stage = 0, implicit = 0, tcur = 0
[DEBUG][rank 0][arkStep_TakeStep_Z][explicit stage] z_0(:) =
 1.224744871391589e+00
 1.732050807568877e+00
\end{lstlisting}

\section{Numerical Experiments}
\label{sec:results}

In this section, we present numerical experiments to test the convergence and performance properties of a sample of the newly implemented methods.
\rev{Numerical experiments with the new multirate adaptivity features are presented in \cite{reynolds_efficient_2025} and 
results utilizing the new Anderson acceleration features are discussed in \cite{gardner2024towards}.
Additional examples are also available in the \code{examples} directory distributed with SUNDIALS.}

\rev{The source code for the experiments below is available at \url{https://github.com/sundials-codes/release-7.4.0-experiments}.
All experiments were conducted on the Dane computing cluster at Lawrence Livermore National Laboratory. Each Dane node consists of two Intel Sapphire Rapids CPUs with a total of 112 cores, 256 GiB of DDR5 memory, and runs the TOSS4 operating system. All the experiments were compiled with GCC 8.5.0.}
%While we do not show multirate adaptivity experiments here, an extensive evaluation can be found in a forthcoming publication.

\subsection{Gray--Scott Diffusion Reaction PDE}

We consider the following PDE from \cite[eq. 2]{pearson1993complex}:
\begin{alignat*}{2}
    \pdv{u}{t} &= \epsilon_1 \nabla^2 u - u v^2 + a (1 - u),
    &\qquad
    \pdv{v}{t} &= \epsilon_2 \nabla^2 v + u v^2 - (a + b) v, \\
    u(0, x, y) &= 1 - \exp\mleft(-80 \mleft((x + 0.05)^2 + (y + 0.02)^2\mright)\mright),
    & \qquad
    v(0, x, y) &= \exp\mleft(-80 \mleft((x - 0.05)^2 + (y - 0.02)^2\mright)\mright).
\end{alignat*}
We take $\epsilon_1 = 2 \times 10^{-5}$, $\epsilon_2 = 10^{-5}$, $a = 0.04$, $b = 0.06$, and a time span of $[0, 3500]$.
The 2D spatial domain is periodic on $[-1, 1]^2$ and is discretized with second order central finite differences.
This approach yields the semidiscretized form
\begin{equation} \label{eq:Gray-Scott_discretized}
    \dv{t}
    \begin{bmatrix}
        \vec{u}(t) \\ \vec{v}(t)
    \end{bmatrix}
    =
    \underbrace{
        \begin{bmatrix}
            \epsilon_1 D \vec{u}(t) \\ \epsilon_2 D \vec{v}(t)
        \end{bmatrix}
    }_{f_3}
    +
    \underbrace{
        \begin{bmatrix}
            0 \\
            \vec{u}(t) \odot \vec{v}(t) \odot \vec{v}(t) - (a + b) \vec{v}(t)
        \end{bmatrix}
    }_{f_2}
    +
    \underbrace{
        \begin{bmatrix}
            -\vec{u}(t) \odot \vec{v}(t) \odot \vec{v}(t) + a (1 - \vec{u}(t)) \\
            0
        \end{bmatrix}
    }_{f_1},
\end{equation}
where $D$ is a discrete Laplacian matrix and $\odot$ denotes the element-wise product of vectors.
The software implementation uses OpenMP to parallelize right-hand side evaluations and internal SUNDIALS computations over an arbitrary number of threads.
For the experiments, we use 50 threads and generate reference solutions with a fifth order explicit Runge--Kutta method using relative and absolute tolerances of $10^{-13}$.

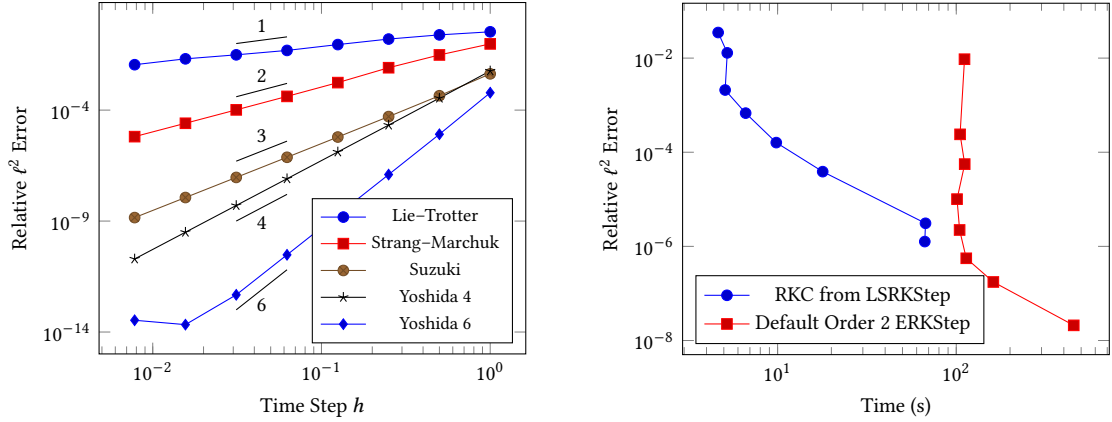
\begin{figure}[ht!]
    \centering
    \begin{subfigure}[t]{0.48\textwidth}
        \begin{tikzpicture}
            \begin{loglogaxis}[
                width=\linewidth,
                xlabel={Time Step $h$},
                ylabel={Relative $\ell^2$ Error},
                % log basis x=2,
                % xmax=3,
                legend style={font=\footnotesize},
                legend entries={Lie--Trotter,Strang--Marchuk,Suzuki,Yoshida 4,Yoshida 6},
                legend pos=south east
            ]
            \foreach \n in {1, 2, 3, 4, 6} {
                \addplot table[x=h, y={Order \n\space Error}]{splitting.csv};
            }
            \draw (axis cs:2^-5,1e-1) -- node[above]{\small 1} (axis cs:2^-4,2e-1);
            \draw (axis cs:2^-5,4e-4) -- node[above]{\small 2} (axis cs:2^-4,1.6e-3);
            \draw (axis cs:2^-5,5e-7) -- node[above]{\small 3} (axis cs:2^-4,4e-6);
            \draw (axis cs:2^-5,1e-9) -- node[below]{\small 4} (axis cs:2^-4,1.6e-8);
            \draw (axis cs:2^-5,1e-13) -- node[below]{\small 6} (axis cs:2^-4,6.4e-12);
            \end{loglogaxis}
        \end{tikzpicture}
        \caption{The default operator splitting methods up to order six implemented in SplittingStep converge at the expected orders. This convergence study uses fixed time steps in the range $2^{-i}$ for $i = 0, \dots, 7$.}
        \label{fig:Gray-Scott:Splitting}
    \end{subfigure}
    \hfill
    \begin{subfigure}[t]{0.48\textwidth}
        \begin{tikzpicture}
            \begin{loglogaxis}[
                width=\linewidth,
                xlabel={Time (s)},
                ylabel={Relative $\ell^2$ Error},
                legend entries={RKC from LSRKStep, Default Order 2 ERKStep},
                legend pos=south west
            ]
            \addplot table[x=LSRK Time,y=LSRK Error]{lsrk.csv};
            \addplot table[x=ERK Time,y=ERK Error]{lsrk.csv};
            % \addplot table[x=DIRK Time,y=DIRK Error]{lsrk.csv};
            % \addplot table[x=DIRK Time,y=DIRK Error]{lsrk_gmres.csv};
            \end{loglogaxis}
        \end{tikzpicture}
        \caption{At loose tolerances, the RKC method implemented in LSRKStep provides an order of magnitude speedup over the default second order explicit Runge--Kutta methods in ARKODE. All methods were run with adaptive time steps, and the data points correspond relative tolerances of $10^{-i}$ for $i = 2, \dots, 9$ and an absolute tolerance of $10^{-13}$.}
        \label{fig:Gray-Scott:LSRK}
    \end{subfigure}
    \caption{Results from applying the operator splitting methods (left) and a low storage Runge--Kutta method (right) to the Gray--Scott problem in \cref{eq:Gray-Scott_discretized}.}
    \label{fig:Gray-Scott}
\end{figure}

As the grouping of terms in \cref{eq:Gray-Scott_discretized} suggests, we use SplittingStep to solve this system with $P = 3$ \rev{operators}.
\rev{Operator} one is a set of scalar, linear ODEs, and \rev{operator} two is a set of scalar, Riccati ODEs.
We create custom \code{SUNStepper}s to \rev{integrate those operators} analytically.
\rev{Operator three contains the diffusion terms and is treated with an explicit Runge--Kutta method of the same order as the operator splitting scheme.}
\rev{We use a $128 \times 128$ grid in space and the default SplittingStep methods of order one to four and six in time.
Respectively, this corresponds to the Lie--Trotter, Strang--Marchuk \cite{strang1968construction,Marchuk1968}, Suzuki \cite{suzuki1992general}, fourth order Yoshida \cite{yoshida1990construction}, and sixth order Yoshida \cite{yoshida1990construction} methods, all of which are suitable for problems with $P=3$ (or any number of) operators.}
\Cref{fig:Gray-Scott:Splitting} confirms the methods converge at their expected orders.

In addition, we compare the performance of the RKC method from LSRKStep with the default second order explicit Runge--Kutta method implemented in the ARKODE ERKStep module on the \rev{unsplit} problem.
We use a finer $1024 \times 1024$ spatial grid to produce a moderately stiff system of ODEs.
\Cref{fig:Gray-Scott:LSRK} shows RKC can attain over an order of magnitude speedup at the loosest tolerances where the number of RKC stages is as large as 42.
The default second order ERKStep method has just three stages, and its limited stability causes the runtime to stagnate at around 100 seconds.

% \subsection{SPRKStep Kepler Experiment?}

\subsection{Adjoint Sensitivity Analysis Applied to a Predator-Prey Model}

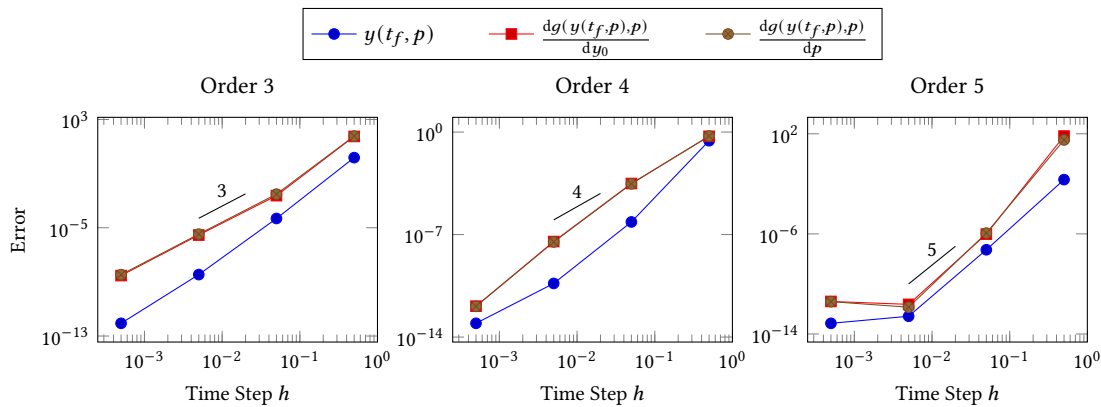
\begin{figure}
    \centering
    \begin{tikzpicture}
        \begin{groupplot}[
            group style={group size=3 by 1, ylabels at=edge left},
            xmode=log,
            ymode=log,
            width=5.3cm,
            xlabel={Time Step $h$},
            ylabel={Error},
            legend columns=3,
            legend style={/tikz/every even column/.append style={column sep=2em}}]
        ]
            \nextgroupplot[
                title={Order 3},
                legend entries={{$y(t_f, p)$}, $\dv{g(y(t_f,p),p)}{y_0}$, $\dv{g(y(t_f,p),p)}{p}$},
                legend to name=leg:asa
            ]
            \foreach \n in {1, 2, 3} {
                \addplot table[x=h,y index=\n]{error_asa_order_3.csv};
            }
            \coordinate (c1) at (rel axis cs:0,1);
            \draw (axis cs:5e-3,5e-5) -- node[above]{\small 3} (axis cs:2e-2,32e-4);
            
            \nextgroupplot[title={Order 4}]
            \foreach \n in {1, 2, 3} {
                \addplot table[x=h,y index=\n]{error_asa_order_4.csv};
            }
            \draw (axis cs:5e-3,1e-6) -- node[above]{\small 4} (axis cs:2e-2,64e-6);
            
            \nextgroupplot[title={Order 5}]
            \foreach \n in {1, 2, 3} {
                \addplot table[x=h,y index=\n]{error_asa_order_5.csv};
            }
            \coordinate (c2) at (rel axis cs:1,1);
            \draw (axis cs:5e-3,1e-10) -- node[above]{\small 5} (axis cs:2e-2,1024e-10);
            % \draw (axis cs:5e-3,1e-9) -- node[below]{\small 4} (axis cs:2e-2,256e-9);
        \end{groupplot}
        \coordinate (c3) at ($(c1)!.5!(c2)$);
        \node[above] at (c3 |- current bounding box.north) {\pgfplotslegendfromname{leg:asa}};
    \end{tikzpicture}
    \caption{Results from the new discrete adjoint sensitivity analysis capability in ARKODE. The methods converge at the expected order for the forward solution as well as the adjoint solution.}
    \label{fig:asa_convergence}
\end{figure}

To demonstrate the convergence of the new discrete ASA implementation in ARKODE, we consider the Lotka--Volterra predator-prey model,
\begin{equation}
\dot{y}(t, p) = 
\begin{bmatrix}
    \dot{y_1}(t, p) \\
    \dot{y_2}(t, p)
\end{bmatrix}
=
\begin{bmatrix}
    p_0 y_1(t, p) - p_1 y_1(t, p) y_2(t, p) \\
    -p_2 y_2(t, p) + p_3 y_1(t, p) y_2(t, p)
\end{bmatrix},\quad
y(t_0, p) = [1, 1]^T,\quad
p = [1.5, 1.0, 3.0, 1.0],\quad
t \in [0, 10],
\end{equation}
and the cost function,
\begin{equation}\label{eq:lv_cost}
    g(y(t_f, p), p) = \frac{1}{2} \| 1 - y(t_f, p) \|^2.
\end{equation}
Reference solutions are generated by solving the forward IVP with Verner's 9th order method \cite{verner1996high} and tight tolerances through the Julia \rev{(v1.8.3)} OrdinaryDiffEq.jl \cite{DifferentialEquations.jl-2017} package, and obtaining the corresponding adjoint solution with reverse-mode automatic differentiation through the Zygote.jl \cite{Zygote.jl-2018} and SciMLSensitivity.jl packages \cite{rackauckas2020universal}.
We solve the forward problem with the ARKODE ARKStep module with fixed time step sizes of $0.5, 0.05, 0.005, 0.0005$, and employ the \code{SUNAdjointStepper} to solve for the gradients $\dv{g}{y_0}$ and $\dv{g}{p}$.
Every other time step is saved as a checkpoint during the forward integration, and the missing steps are recomputed during the adjoint integration.
To compare the obtained solutions with the reference solutions, we use the error metric, $\text{Error}(x, x_{ref}) = (\|x\|_2 - \|x_{ref}\|_2)/\|x_{ref}\|_2$, where $x$ is one of $y(t_f, p)$, $\dv{g(y(t_f,p),p)}{y_0}$, and $\dv{g(y(t_f,p),p)}{p}$, and $x_{ref}$ is the corresponding reference solution at $t_f$.
The comparison (\Cref{fig:asa_convergence}) shows that the methods converge at the expected order for the forward solution as well as the adjoint solution.

\section{Conclusions}
\label{sec:conclusions}

This article overviews the new capabilities added to SUNDIALS in versions 6.2.0 (Aug. 2022) through 7.4.0 (June 2025).  These capabilities include additions to the ARKODE package, including low storage Runge--Kutta methods, symplectic partitioned Runge--Kutta methods, low to high order operator splitting methods, multirate temporal adaptivity, and adjoint sensitivity analysis for explicit Runge--Kutta methods, as well as extensions of the Anderson accelerated fixed point solver, and new error handling and logging options throughout all six SUNDIALS packages.
Selected capabilities were demonstrated in \cref{sec:results} showing the expected convergence behaviors.

\begin{acks}
This material is based upon work supported by the U.S. Department of Energy, Office of Science, Office of Advanced Scientific Computing Research (ASCR) via the Frameworks, Algorithms (FASTMath), and \rev{Scalable} Technologies for Mathematics Institute, the Next-Generation Scientific Software Technologies (NGSST) Program, the FRONTIERS in Leadership Gyrokinetic Simulation U. S. Department of Energy Office of Science Office of Fusion Energy Sciences (FES) and ASCR partnership, the CEDA: Computational Evaluation and Design of Actuators for Core-Edge Integration FES and ASCR partnership, and the Traversing the “death valley” separating short and long times in non-equilibrium quantum dynamical simulations of real materials U.S. Department of Energy Office of Science Office of Basic Energy Sciences and ASCR Partnership, all through the Scientific Discovery through Advanced Computing (SciDAC) program.
In addition, some funding for this work was provided by the Lawrence Livermore National Laboratory Institutional Scientific Capability Program.

This work was partially performed under the auspices of the U.S. Department of Energy by Lawrence Livermore National Laboratory under Contract DE-AC52-07NA27344. LLNL-JRNL-2007253.

% This manuscript	has	been authored in part by Lawrence Livermore	National Security, LLC under Contract No. DE-AC52-07NA27344 with the US. Department of Energy. The United States Government retains, and the publisher, by accepting the article for publication, acknowledges that the United States Government retains a non-exclusive, paid-up, irrevocable, world-wide license to publish or reproduce the published form of this manuscript, or allow others to do so, for United States Government purposes.	
\end{acks}

\bibliographystyle{ACM-Reference-Format}
\bibliography{main}


\begin{thebibliography}{51}

%%% the Web address.

\ifx \showCODEN    \undefined \def \showCODEN     #1{\unskip}     \fi
\ifx \showISBNx    \undefined \def \showISBNx     #1{\unskip}     \fi
\ifx \showISBNxiii \undefined \def \showISBNxiii  #1{\unskip}     \fi
\ifx \showISSN     \undefined \def \showISSN      #1{\unskip}     \fi
\ifx \showLCCN     \undefined \def \showLCCN      #1{\unskip}     \fi
\ifx \shownote     \undefined \def \shownote      #1{#1}          \fi
\ifx \showarticletitle \undefined \def \showarticletitle #1{#1}   \fi
\ifx \showURL      \undefined \def \showURL       {\relax}        \fi
% invisible to TeX
\providecommand\bibfield[2]{#2}
\providecommand\bibinfo[2]{#2}
\providecommand\natexlab[1]{#1}
\providecommand\showeprint[2][]{arXiv:#2}

\bibitem[Anderson(1965)]%
        {anderson1965iterative}
\bibfield{author}{\bibinfo{person}{Donald~G. Anderson}.}
  \bibinfo{year}{1965}\natexlab{}.
\newblock \showarticletitle{Iterative procedures for nonlinear integral
  equations}.
\newblock \bibinfo{journal}{\emph{J. ACM}} \bibinfo{volume}{12},
  \bibinfo{number}{4} (\bibinfo{year}{1965}), \bibinfo{pages}{547--560}.
\newblock
\href{https://doi.org/10.1145/321296.321305}{doi:\nolinkurl{10.1145/321296.321305}}


\bibitem[Balos et~al\mbox{.}(2025)]%
        {balos_sundials_2025}
\bibfield{author}{\bibinfo{person}{Cody~J. Balos}, \bibinfo{person}{Marcus
  Day}, \bibinfo{person}{Lucas Esclapez}, \bibinfo{person}{Anne~M. Felden},
  \bibinfo{person}{David~J. Gardner}, \bibinfo{person}{Malik Hassanaly},
  \bibinfo{person}{Daniel~R. Reynolds}, \bibinfo{person}{Jon~S. Rood},
  \bibinfo{person}{Jean~M. Sexton}, \bibinfo{person}{Nicholas~T. Wimer}, {and}
  \bibinfo{person}{Carol~S. Woodward}.} \bibinfo{year}{2025}\natexlab{}.
\newblock \showarticletitle{{SUNDIALS} time integrators for exascale
  applications with many independent systems of ordinary differential
  equations}.
\newblock \bibinfo{journal}{\emph{The International Journal of High Performance
  Computing Applications}} \bibinfo{volume}{39}, \bibinfo{number}{1}
  (\bibinfo{date}{Jan.} \bibinfo{year}{2025}), \bibinfo{pages}{123--146}.
\newblock
\showISSN{1094-3420}
\href{https://doi.org/10.1177/10943420241280060}{doi:\nolinkurl{10.1177/10943420241280060}}
\newblock
\shownote{Publisher: SAGE Publications Ltd STM}.


\bibitem[Blanes et~al\mbox{.}(2024)]%
        {blanes2024splitting}
\bibfield{author}{\bibinfo{person}{Sergio Blanes}, \bibinfo{person}{Fernando
  Casas}, {and} \bibinfo{person}{Ander Murua}.}
  \bibinfo{year}{2024}\natexlab{}.
\newblock \showarticletitle{Splitting methods for differential equations}.
\newblock \bibinfo{journal}{\emph{Acta Numerica}}  \bibinfo{volume}{33}
  (\bibinfo{year}{2024}), \bibinfo{pages}{1–161}.
\newblock
\href{https://doi.org/10.1017/S0962492923000077}{doi:\nolinkurl{10.1017/S0962492923000077}}


\bibitem[Chen and Vuik(2024)]%
        {chen2024non}
\bibfield{author}{\bibinfo{person}{Kewang Chen} {and} \bibinfo{person}{Cornelis
  Vuik}.} \bibinfo{year}{2024}\natexlab{}.
\newblock \showarticletitle{Non-stationary {A}nderson acceleration with
  optimized damping}.
\newblock \bibinfo{journal}{\emph{J. Comput. Appl. Math.}}
  \bibinfo{volume}{451} (\bibinfo{year}{2024}), \bibinfo{pages}{116077}.
\newblock
\showISSN{0377-0427}
\href{https://doi.org/10.1016/j.cam.2024.116077}{doi:\nolinkurl{10.1016/j.cam.2024.116077}}


\bibitem[Creutz and Gocksch(1989)]%
        {CrGo:89}
\bibfield{author}{\bibinfo{person}{Michael Creutz} {and}
  \bibinfo{person}{Andreas Gocksch}.} \bibinfo{year}{1989}\natexlab{}.
\newblock \showarticletitle{Higher-order hybrid {M}onte {C}arlo algorithms}.
\newblock \bibinfo{journal}{\emph{Phys. Rev. Lett.}}  \bibinfo{volume}{63}
  (\bibinfo{date}{Jul} \bibinfo{year}{1989}), \bibinfo{pages}{9--12}.
\newblock
Issue 1.
\href{https://doi.org/10.1103/PhysRevLett.63.9}{doi:\nolinkurl{10.1103/PhysRevLett.63.9}}


\bibitem[Donahue and Caldwell(2018)]%
        {donahue2018impact}
\bibfield{author}{\bibinfo{person}{Aaron~S. Donahue} {and}
  \bibinfo{person}{Peter~M. Caldwell}.} \bibinfo{year}{2018}\natexlab{}.
\newblock \showarticletitle{Impact of Physics Parameterization Ordering in a
  Global Atmosphere Model}.
\newblock \bibinfo{journal}{\emph{Journal of Advances in Modeling Earth
  Systems}} \bibinfo{volume}{10}, \bibinfo{number}{2} (\bibinfo{year}{2018}),
  \bibinfo{pages}{481--499}.
\newblock
\href{https://doi.org/10.1002/2017MS001067}{doi:\nolinkurl{10.1002/2017MS001067}}


\bibitem[Evans et~al\mbox{.}(2020)]%
        {evans2020proof}
\bibfield{author}{\bibinfo{person}{Claire Evans}, \bibinfo{person}{Sara
  Pollock}, \bibinfo{person}{Leo~G. Rebholz}, {and} \bibinfo{person}{Mengying
  Xiao}.} \bibinfo{year}{2020}\natexlab{}.
\newblock \showarticletitle{A proof that {A}nderson acceleration improves the
  convergence rate in linearly converging fixed-point methods (but not in those
  converging quadratically)}.
\newblock \bibinfo{journal}{\emph{SIAM J. Numer. Anal.}} \bibinfo{volume}{58},
  \bibinfo{number}{1} (\bibinfo{year}{2020}), \bibinfo{pages}{788--810}.
\newblock
\href{https://doi.org/10.1137/19M1245384}{doi:\nolinkurl{10.1137/19M1245384}}


\bibitem[Fekete et~al\mbox{.}(2022)]%
        {CondeSSPEmbeddings}
\bibfield{author}{\bibinfo{person}{Imre Fekete}, \bibinfo{person}{Sidafa
  Conde}, {and} \bibinfo{person}{John~N. Shadid}.}
  \bibinfo{year}{2022}\natexlab{}.
\newblock \showarticletitle{Embedded pairs for optimal explicit strong
  stability preserving Runge–Kutta methods}.
\newblock \bibinfo{journal}{\emph{J. Comput. Appl. Math.}}
  \bibinfo{volume}{412} (\bibinfo{year}{2022}), \bibinfo{pages}{114325}.
\newblock
\showISSN{0377-0427}
\href{https://doi.org/10.1016/j.cam.2022.114325}{doi:\nolinkurl{10.1016/j.cam.2022.114325}}


\bibitem[Fish et~al\mbox{.}(2024)]%
        {fish_implicitexplicit_2024}
\bibfield{author}{\bibinfo{person}{Alex~C. Fish}, \bibinfo{person}{Daniel~R.
  Reynolds}, {and} \bibinfo{person}{Steven~B. Roberts}.}
  \bibinfo{year}{2024}\natexlab{}.
\newblock \showarticletitle{Implicit–explicit multirate infinitesimal
  stage-restart methods}.
\newblock \bibinfo{journal}{\emph{J. Comput. Appl. Math.}}
  \bibinfo{volume}{438} (\bibinfo{date}{March} \bibinfo{year}{2024}),
  \bibinfo{pages}{115534}.
\newblock
\showISSN{0377-0427}
\href{https://doi.org/10.1016/j.cam.2023.115534}{doi:\nolinkurl{10.1016/j.cam.2023.115534}}


\bibitem[Gardner et~al\mbox{.}(2024)]%
        {gardner2024towards}
\bibfield{author}{\bibinfo{person}{David~J. Gardner}, \bibinfo{person}{Linda
  LoDestro}, {and} \bibinfo{person}{Carol~S. Woodward}.}
  \bibinfo{year}{2024}\natexlab{}.
\newblock \showarticletitle{Towards the Use of {Anderson} Acceleration in
  Coupled Transport-Gyrokinetic Turbulence Simulations}.
\newblock  (\bibinfo{year}{2024}).
\newblock
\showeprint[arxiv]{2407.03561}


\bibitem[Gardner et~al\mbox{.}(2022)]%
        {gardner2022sundials}
\bibfield{author}{\bibinfo{person}{David~J. Gardner},
  \bibinfo{person}{Daniel~R. Reynolds}, \bibinfo{person}{Carol~S. Woodward},
  {and} \bibinfo{person}{Cody~J. Balos}.} \bibinfo{year}{2022}\natexlab{}.
\newblock \showarticletitle{Enabling new flexibility in the {SUNDIALS} suite of
  nonlinear and differential/algebraic equation solvers}.
\newblock \bibinfo{journal}{\emph{ACM Transactions on Mathematical Software
  (TOMS)}} \bibinfo{volume}{48}, \bibinfo{number}{3} (\bibinfo{year}{2022}),
  \bibinfo{pages}{1--24}.
\newblock
\href{https://doi.org/10.1145/3539801}{doi:\nolinkurl{10.1145/3539801}}


\bibitem[Gholami et~al\mbox{.}(2019)]%
        {gholami2019anode}
\bibfield{author}{\bibinfo{person}{Amir Gholami}, \bibinfo{person}{Kurt
  Keutzer}, {and} \bibinfo{person}{George Biros}.}
  \bibinfo{year}{2019}\natexlab{}.
\newblock \showarticletitle{{ANODE}: unconditionally accurate memory-efficient
  gradients for neural {ODE}s}. In \bibinfo{booktitle}{\emph{Proceedings of the
  28th International Joint Conference on Artificial Intelligence}}
  \emph{(\bibinfo{series}{IJCAI'19})}. \bibinfo{publisher}{AAAI Press},
  \bibinfo{address}{Macao, China}, \bibinfo{pages}{730–736}.
\newblock
\showISBNx{9780999241141}


\bibitem[Giles and Pierce(2000)]%
        {giles2000introduction}
\bibfield{author}{\bibinfo{person}{Michael~B. Giles} {and}
  \bibinfo{person}{Niles~A. Pierce}.} \bibinfo{year}{2000}\natexlab{}.
\newblock \showarticletitle{An introduction to the adjoint approach to design}.
\newblock \bibinfo{journal}{\emph{Flow, turbulence and combustion}}
  \bibinfo{volume}{65}, \bibinfo{number}{3} (\bibinfo{year}{2000}),
  \bibinfo{pages}{393--415}.
\newblock


\bibitem[Guenter et~al\mbox{.}(2024)]%
        {guenter2024pythos}
\bibfield{author}{\bibinfo{person}{Victoria Guenter}, \bibinfo{person}{Siqi
  Wei}, {and} \bibinfo{person}{Raymond~J. Spiteri}.}
  \bibinfo{year}{2024}\natexlab{}.
\newblock \showarticletitle{{pythOS}: {A} Python library for solving {IVP}s by
  operator splitting}.
\newblock \bibinfo{journal}{\emph{arXiv preprint arXiv:2407.05475}}
  (\bibinfo{year}{2024}).
\newblock


\bibitem[Hager(2000)]%
        {hager2000runge}
\bibfield{author}{\bibinfo{person}{William~W Hager}.}
  \bibinfo{year}{2000}\natexlab{}.
\newblock \showarticletitle{{Runge-Kutta} methods in optimal control and the
  transformed adjoint system}.
\newblock \bibinfo{journal}{\emph{Numer. Math.}}  \bibinfo{volume}{87}
  (\bibinfo{year}{2000}), \bibinfo{pages}{247--282}.
\newblock
\href{https://doi.org/10.1007/s002110000178}{doi:\nolinkurl{10.1007/s002110000178}}


\bibitem[Hairer et~al\mbox{.}(2006)]%
        {hairerWarner2006geometric}
\bibfield{author}{\bibinfo{person}{Ernst Hairer}, \bibinfo{person}{Gerhard
  Wanner}, {and} \bibinfo{person}{Christian Lubich}.}
  \bibinfo{year}{2006}\natexlab{}.
\newblock \bibinfo{booktitle}{\emph{Geometric Numerical Integration,
  Structure-Preserving Algorithms for Ordinary Differential Equations}}.
\newblock \bibinfo{publisher}{Springer Series in Computational Mathematics}.
\newblock
\href{https://doi.org/10.1007/3-540-30666-8}{doi:\nolinkurl{10.1007/3-540-30666-8}}


\bibitem[Hindmarsh et~al\mbox{.}(2005)]%
        {hindmarsh2005sundials}
\bibfield{author}{\bibinfo{person}{Alan~C. Hindmarsh},
  \bibinfo{person}{Peter~N. Brown}, \bibinfo{person}{Keith~E. Grant},
  \bibinfo{person}{Steven~L. Lee}, \bibinfo{person}{Radu Serban},
  \bibinfo{person}{Dan~E. Shumaker}, {and} \bibinfo{person}{Carol~S.
  Woodward}.} \bibinfo{year}{2005}\natexlab{}.
\newblock \showarticletitle{{SUNDIALS}: Suite of nonlinear and
  differential/algebraic equation solvers}.
\newblock \bibinfo{journal}{\emph{ACM Transactions on Mathematical Software
  (TOMS)}} \bibinfo{volume}{31}, \bibinfo{number}{3} (\bibinfo{year}{2005}),
  \bibinfo{pages}{363--396}.
\newblock
\href{https://doi.org/10.1145/1089014.1089020}{doi:\nolinkurl{10.1145/1089014.1089020}}


\bibitem[Hindmarsh et~al\mbox{.}(2025)]%
        {kinsolDocumentation}
\bibfield{author}{\bibinfo{person}{Alan~C. Hindmarsh}, \bibinfo{person}{Radu
  Serban}, \bibinfo{person}{Cody~J. Balos}, \bibinfo{person}{David~J. Gardner},
  \bibinfo{person}{Daniel~R. Reynolds}, {and} \bibinfo{person}{Carol~S.
  Woodward}.} \bibinfo{year}{2025}\natexlab{}.
\newblock \bibinfo{title}{User Documentation for KINSOL}.
\newblock
  \bibinfo{howpublished}{url{https://sundials.readthedocs.io/en/latest/kinsol}}.
\newblock
\urldef\tempurl%
\url{https://sundials.readthedocs.io/en/latest/kinsol}
\showURL{%
\tempurl}
\newblock
\shownote{v7.3.0}.


\bibitem[Innes(2018)]%
        {Zygote.jl-2018}
\bibfield{author}{\bibinfo{person}{Michael Innes}.}
  \bibinfo{year}{2018}\natexlab{}.
\newblock \showarticletitle{Don't Unroll Adjoint: Differentiating SSA-Form
  Programs}.
\newblock \bibinfo{journal}{\emph{CoRR}}  \bibinfo{volume}{abs/1810.07951}
  (\bibinfo{year}{2018}).
\newblock
\showeprint[arxiv]{1810.07951}
\urldef\tempurl%
\url{http://arxiv.org/abs/1810.07951}
\showURL{%
\tempurl}


\bibitem[Ketcheson(2008)]%
        {KetchesonHighlyEfficientSSP}
\bibfield{author}{\bibinfo{person}{David~I. Ketcheson}.}
  \bibinfo{year}{2008}\natexlab{}.
\newblock \showarticletitle{Highly Efficient Strong Stability-Preserving
  Runge–Kutta Methods with Low-Storage Implementations}.
\newblock \bibinfo{journal}{\emph{SIAM Journal on Scientific Computing}}
  \bibinfo{volume}{30}, \bibinfo{number}{4} (\bibinfo{year}{2008}),
  \bibinfo{pages}{2113--2136}.
\newblock
\href{https://doi.org/10.1137/07070485X}{doi:\nolinkurl{10.1137/07070485X}}


\bibitem[Lockhart et~al\mbox{.}(2022)]%
        {lockhart2022performance}
\bibfield{author}{\bibinfo{person}{Shelby Lockhart}, \bibinfo{person}{David~J.
  Gardner}, \bibinfo{person}{Carol~S. Woodward}, \bibinfo{person}{Stephen
  Thomas}, {and} \bibinfo{person}{Luke~N. Olson}.}
  \bibinfo{year}{2022}\natexlab{}.
\newblock \showarticletitle{Performance of low synchronization
  orthogonalization methods in {A}nderson accelerated fixed point solvers}. In
  \bibinfo{booktitle}{\emph{Proceedings of the 2022 SIAM Conference on Parallel
  Processing for Scientific Computing}}. SIAM, \bibinfo{publisher}{Society for
  Industrial and Applied Mathematics}, \bibinfo{pages}{49--59}.
\newblock
\href{https://doi.org/10.1137/1.9781611977141.5}{doi:\nolinkurl{10.1137/1.9781611977141.5}}


\bibitem[Luan et~al\mbox{.}(2020)]%
        {luan_new_2020}
\bibfield{author}{\bibinfo{person}{Vu~Thai Luan}, \bibinfo{person}{Rujeko
  Chinomona}, {and} \bibinfo{person}{Daniel~R. Reynolds}.}
  \bibinfo{year}{2020}\natexlab{}.
\newblock \showarticletitle{A {New} {Class} of {High}-{Order} {Methods} for
  {Multirate} {Differential} {Equations}}.
\newblock \bibinfo{journal}{\emph{SIAM J. Sci. Comput.}} \bibinfo{volume}{42},
  \bibinfo{number}{2} (\bibinfo{date}{Jan.} \bibinfo{year}{2020}),
  \bibinfo{pages}{A1245--A1268}.
\newblock
\showISSN{1064-8275, 1095-7197}
\href{https://doi.org/10.1137/19M125621X}{doi:\nolinkurl{10.1137/19M125621X}}


\bibitem[Marchuk(1968)]%
        {Marchuk1968}
\bibfield{author}{\bibinfo{person}{Gurij~Ivanovich Marchuk}.}
  \bibinfo{year}{1968}\natexlab{}.
\newblock \showarticletitle{Some application of splitting-up methods to the
  solution of mathematical physics problems}.
\newblock \bibinfo{journal}{\emph{Aplikace Matematiky}} \bibinfo{volume}{13},
  \bibinfo{number}{2} (\bibinfo{year}{1968}), \bibinfo{pages}{103--132}.
\newblock
\urldef\tempurl%
\url{http://eudml.org/doc/14518}
\showURL{%
\tempurl}


\bibitem[McLachlan and Quispel(2002)]%
        {mclachlan2002splitting}
\bibfield{author}{\bibinfo{person}{Robert~I. McLachlan} {and}
  \bibinfo{person}{G.~Reinout~W. Quispel}.} \bibinfo{year}{2002}\natexlab{}.
\newblock \showarticletitle{Splitting methods}.
\newblock \bibinfo{journal}{\emph{Acta Numerica}}  \bibinfo{volume}{11}
  (\bibinfo{year}{2002}), \bibinfo{pages}{341–434}.
\newblock
\href{https://doi.org/10.1017/S0962492902000053}{doi:\nolinkurl{10.1017/S0962492902000053}}


\bibitem[Meyer et~al\mbox{.}(2012)]%
        {meyerSecondorderAccurateSuper2012}
\bibfield{author}{\bibinfo{person}{Chad~D. Meyer}, \bibinfo{person}{Dinshaw~S.
  Balsara}, {and} \bibinfo{person}{Tariq~D. Aslam}.}
  \bibinfo{year}{2012}\natexlab{}.
\newblock \showarticletitle{A Second-Order Accurate {{Super TimeStepping}}
  Formulation for Anisotropic Thermal Conduction}.
\newblock \bibinfo{journal}{\emph{Monthly Notices of the Royal Astronomical
  Society}} \bibinfo{volume}{422}, \bibinfo{number}{3} (\bibinfo{date}{May}
  \bibinfo{year}{2012}), \bibinfo{pages}{2102--2115}.
\newblock
\showISSN{00358711}
\href{https://doi.org/10.1111/j.1365-2966.2012.20744.x}{doi:\nolinkurl{10.1111/j.1365-2966.2012.20744.x}}


\bibitem[Meyer et~al\mbox{.}(2014)]%
        {meyerStabilizedRungeKutta2014a}
\bibfield{author}{\bibinfo{person}{Chad~D. Meyer}, \bibinfo{person}{Dinshaw~S.
  Balsara}, {and} \bibinfo{person}{Tariq~D. Aslam}.}
  \bibinfo{year}{2014}\natexlab{}.
\newblock \showarticletitle{A Stabilized
  {{Runge}}{\textendash}{{Kutta}}{\textendash}{{Legendre}} Method for Explicit
  Super-Time-Stepping of Parabolic and Mixed Equations}.
\newblock \bibinfo{journal}{\emph{J. Comput. Phys.}}  \bibinfo{volume}{257}
  (\bibinfo{date}{Jan.} \bibinfo{year}{2014}), \bibinfo{pages}{594--626}.
\newblock
\showISSN{00219991}
\href{https://doi.org/10.1016/j.jcp.2013.08.021}{doi:\nolinkurl{10.1016/j.jcp.2013.08.021}}


\bibitem[Moayeri and Spiteri(2025)]%
        {moayeri2025tost}
\bibfield{author}{\bibinfo{person}{Mohammad~Mahdi Moayeri} {and}
  \bibinfo{person}{Raymond~J. Spiteri}.} \bibinfo{year}{2025}\natexlab{}.
\newblock \showarticletitle{tost.{II}: A temporal operator-splitting template
  library in deal.{II}}.
\newblock \bibinfo{journal}{\emph{Mathematics and Computers in Simulation}}
  \bibinfo{volume}{241} (\bibinfo{year}{2025}), \bibinfo{pages}{790--804}.
\newblock
\showISSN{0378-4754}
\href{https://doi.org/10.1016/j.matcom.2025.09.005}{doi:\nolinkurl{10.1016/j.matcom.2025.09.005}}


\bibitem[Pearson(1993)]%
        {pearson1993complex}
\bibfield{author}{\bibinfo{person}{John~E. Pearson}.}
  \bibinfo{year}{1993}\natexlab{}.
\newblock \showarticletitle{Complex Patterns in a Simple System}.
\newblock \bibinfo{journal}{\emph{Science}} \bibinfo{volume}{261},
  \bibinfo{number}{5118} (\bibinfo{year}{1993}), \bibinfo{pages}{189--192}.
\newblock
\href{https://doi.org/10.1126/science.261.5118.189}{doi:\nolinkurl{10.1126/science.261.5118.189}}


\bibitem[Pollock and Rebholz(2021)]%
        {pollock2021anderson}
\bibfield{author}{\bibinfo{person}{Sara Pollock} {and} \bibinfo{person}{Leo~G.
  Rebholz}.} \bibinfo{year}{2021}\natexlab{}.
\newblock \showarticletitle{{A}nderson acceleration for contractive and
  noncontractive operators}.
\newblock \bibinfo{journal}{\emph{IMA J. Numer. Anal.}} \bibinfo{volume}{41},
  \bibinfo{number}{4} (\bibinfo{year}{2021}), \bibinfo{pages}{2841--2872}.
\newblock
\href{https://doi.org/10.1093/imanum/draa095}{doi:\nolinkurl{10.1093/imanum/draa095}}


\bibitem[Pollock and Rebholz(2023)]%
        {pollock2023filtering}
\bibfield{author}{\bibinfo{person}{Sara Pollock} {and} \bibinfo{person}{Leo~G.
  Rebholz}.} \bibinfo{year}{2023}\natexlab{}.
\newblock \showarticletitle{Filtering for {A}nderson acceleration}.
\newblock \bibinfo{journal}{\emph{SIAM Journal on Scientific Computing}}
  \bibinfo{volume}{45}, \bibinfo{number}{4} (\bibinfo{year}{2023}),
  \bibinfo{pages}{A1571--A1590}.
\newblock
\href{https://doi.org/10.1137/22M1536741}{doi:\nolinkurl{10.1137/22M1536741}}


\bibitem[Rackauckas et~al\mbox{.}(2020)]%
        {rackauckas2020universal}
\bibfield{author}{\bibinfo{person}{Christopher Rackauckas},
  \bibinfo{person}{Yingbo Ma}, \bibinfo{person}{Julius Martensen},
  \bibinfo{person}{Collin Warner}, \bibinfo{person}{Kirill Zubov},
  \bibinfo{person}{Rohit Supekar}, \bibinfo{person}{Dominic Skinner}, {and}
  \bibinfo{person}{Ali Ramadhan}.} \bibinfo{year}{2020}\natexlab{}.
\newblock \showarticletitle{Universal differential equations for scientific
  machine learning}.
\newblock \bibinfo{journal}{\emph{arXiv preprint arXiv:2001.04385}}
  (\bibinfo{year}{2020}).
\newblock


\bibitem[Rackauckas and Nie(2017)]%
        {DifferentialEquations.jl-2017}
\bibfield{author}{\bibinfo{person}{Christopher Rackauckas} {and}
  \bibinfo{person}{Qing Nie}.} \bibinfo{year}{2017}\natexlab{}.
\newblock \showarticletitle{DifferentialEquations.jl – A Performant and
  Feature-Rich Ecosystem for Solving Differential Equations in Julia}.
\newblock \bibinfo{journal}{\emph{The Journal of Open Research Software}}
  \bibinfo{volume}{5}, \bibinfo{number}{1} (\bibinfo{year}{2017}).
\newblock
\href{https://doi.org/10.5334/jors.151}{doi:\nolinkurl{10.5334/jors.151}}


\bibitem[Reynolds et~al\mbox{.}(2025a)]%
        {reynolds_efficient_2025}
\bibfield{author}{\bibinfo{person}{Daniel~R. Reynolds}, \bibinfo{person}{Sylvia
  Amihere}, \bibinfo{person}{Dashon Mitchell}, {and} \bibinfo{person}{Vu~Thai
  Luan}.} \bibinfo{year}{2025}\natexlab{a}.
\newblock \showarticletitle{Efficient and {Flexible} {Multirate} {Temporal}
  {Adaptivity}}.
\newblock  (\bibinfo{year}{2025}).
\newblock
\href{https://doi.org/10.48550/arXiv.2510.14964}{doi:\nolinkurl{10.48550/arXiv.2510.14964}}
\showeprint[arxiv]{2510.14964}


\bibitem[Reynolds et~al\mbox{.}(2025b)]%
        {arkodeDocumentation}
\bibfield{author}{\bibinfo{person}{Daniel~R. Reynolds},
  \bibinfo{person}{David~J. Gardner}, \bibinfo{person}{Rujeko~Chinomona Carol
  S.~Woodward}, {and} \bibinfo{person}{Cody~J. Balos}.}
  \bibinfo{year}{2025}\natexlab{b}.
\newblock \bibinfo{title}{User Documentation for ARKODE}.
\newblock
  \bibinfo{howpublished}{url{https://sundials.readthedocs.io/en/latest/arkode}}.
\newblock
\urldef\tempurl%
\url{https://sundials.readthedocs.io/en/latest/arkode}
\showURL{%
\tempurl}
\newblock
\shownote{v6.3.0}.


\bibitem[Reynolds et~al\mbox{.}(2023)]%
        {reynolds2023arkode}
\bibfield{author}{\bibinfo{person}{Daniel~R. Reynolds},
  \bibinfo{person}{David~J. Gardner}, \bibinfo{person}{Carol~S. Woodward},
  {and} \bibinfo{person}{Rujeko Chinomona}.} \bibinfo{year}{2023}\natexlab{}.
\newblock \showarticletitle{{ARKODE: A flexible IVP solver infrastructure for
  one-step methods}}.
\newblock \bibinfo{journal}{\emph{ACM Trans. Math. Software}}
  \bibinfo{volume}{49}, \bibinfo{number}{2} (\bibinfo{year}{2023}),
  \bibinfo{pages}{1--26}.
\newblock
\href{https://doi.org/10.1145/3594632}{doi:\nolinkurl{10.1145/3594632}}


\bibitem[Roberts et~al\mbox{.}(2022)]%
        {roberts_fast_2022}
\bibfield{author}{\bibinfo{person}{Steven Roberts}, \bibinfo{person}{Andrey~A.
  Popov}, \bibinfo{person}{Arash Sarshar}, {and} \bibinfo{person}{Adrian
  Sandu}.} \bibinfo{year}{2022}\natexlab{}.
\newblock \showarticletitle{A {Fast} {Time}-{Stepping} {Strategy} for
  {Dynamical} {Systems} {Equipped} with a {Surrogate} {Model}}.
\newblock \bibinfo{journal}{\emph{SIAM J. Sci. Comput.}} \bibinfo{volume}{44},
  \bibinfo{number}{3} (\bibinfo{date}{June} \bibinfo{year}{2022}),
  \bibinfo{pages}{A1405--A1427}.
\newblock
\showISSN{1064-8275}
\href{https://doi.org/10.1137/20M1386281}{doi:\nolinkurl{10.1137/20M1386281}}
\newblock
\shownote{Publisher: Society for Industrial and Applied Mathematics}.


\bibitem[Sandu(2006)]%
        {sanduDiscrete2006}
\bibfield{author}{\bibinfo{person}{Adrian Sandu}.}
  \bibinfo{year}{2006}\natexlab{}.
\newblock \showarticletitle{On the Properties of Runge-Kutta Discrete
  Adjoints}. In \bibinfo{booktitle}{\emph{Computational Science -- ICCS 2006}},
  \bibfield{editor}{\bibinfo{person}{Vassil~N. Alexandrov},
  \bibinfo{person}{Geert~Dick van Albada}, \bibinfo{person}{Peter M.~A. Sloot},
  {and} \bibinfo{person}{Jack Dongarra}} (Eds.). \bibinfo{publisher}{Springer
  Berlin Heidelberg}, \bibinfo{address}{Berlin, Heidelberg},
  \bibinfo{pages}{550--557}.
\newblock
\showISBNx{978-3-540-34386-8}
\href{https://doi.org/10.1007/11758549_76}{doi:\nolinkurl{10.1007/11758549_76}}


\bibitem[Sandu(2019)]%
        {sandu_class_2019}
\bibfield{author}{\bibinfo{person}{Adrian Sandu}.}
  \bibinfo{year}{2019}\natexlab{}.
\newblock \showarticletitle{A {Class} of {Multirate} {Infinitesimal} {GARK}
  {Methods}}.
\newblock \bibinfo{journal}{\emph{SIAM J. Numer. Anal.}} \bibinfo{volume}{57},
  \bibinfo{number}{5} (\bibinfo{date}{Jan.} \bibinfo{year}{2019}),
  \bibinfo{pages}{2300--2327}.
\newblock
\showISSN{0036-1429, 1095-7170}
\href{https://doi.org/10.1137/18M1205492}{doi:\nolinkurl{10.1137/18M1205492}}


\bibitem[Sanz-Serna(2016)]%
        {sanzserna2016symplectic}
\bibfield{author}{\bibinfo{person}{J.~M. Sanz-Serna}.}
  \bibinfo{year}{2016}\natexlab{}.
\newblock \showarticletitle{Symplectic Runge--Kutta Schemes for Adjoint
  Equations, Automatic Differentiation, Optimal Control, and More}.
\newblock \bibinfo{journal}{\emph{SIAM Rev.}} \bibinfo{volume}{58},
  \bibinfo{number}{1} (\bibinfo{year}{2016}), \bibinfo{pages}{3--33}.
\newblock
\href{https://doi.org/10.1137/151002769}{doi:\nolinkurl{10.1137/151002769}}


\bibitem[Serban and Hindmarsh(2005)]%
        {serban2005cvodes}
\bibfield{author}{\bibinfo{person}{Radu Serban} {and} \bibinfo{person}{Alan~C.
  Hindmarsh}.} \bibinfo{year}{2005}\natexlab{}.
\newblock \showarticletitle{CVODES: The Sensitivity-Enabled ODE Solver in
  SUNDIALS}.
\newblock   \bibinfo{volume}{Volume 6: 5th International Conference on
  Multibody Systems, Nonlinear Dynamics, and Control, Parts A, B, and C}
  (\bibinfo{date}{09} \bibinfo{year}{2005}), \bibinfo{pages}{257--269}.
\newblock
\href{https://doi.org/10.1115/DETC2005-85597}{doi:\nolinkurl{10.1115/DETC2005-85597}}


\bibitem[Sofroniou and Spaletta(2003)]%
        {sofroniou2003increment}
\bibfield{author}{\bibinfo{person}{Mark Sofroniou} {and}
  \bibinfo{person}{Giulia Spaletta}.} \bibinfo{year}{2003}\natexlab{}.
\newblock \showarticletitle{Increment formulations for rounding error reduction
  in the numerical solution of structured differential systems}.
\newblock \bibinfo{journal}{\emph{Future Generation Computer Systems}}
  \bibinfo{volume}{19}, \bibinfo{number}{3} (\bibinfo{year}{2003}),
  \bibinfo{pages}{375--383}.
\newblock
\showISSN{0167-739X}
\href{https://doi.org/10.1016/S0167-739X(02)00164-4}{doi:\nolinkurl{10.1016/S0167-739X(02)00164-4}}
\newblock
\shownote{Special Issue on Geometric Numerical Algorithms}.


\bibitem[Sommeijer et~al\mbox{.}(1998)]%
        {sommeijer_rkc_1998}
\bibfield{author}{\bibinfo{person}{B.~P. Sommeijer}, \bibinfo{person}{L.~F.
  Shampine}, {and} \bibinfo{person}{J.~G. Verwer}.}
  \bibinfo{year}{1998}\natexlab{}.
\newblock \showarticletitle{{RKC}: {An} explicit solver for parabolic {PDEs}}.
\newblock \bibinfo{journal}{\emph{J. Comput. Appl. Math.}}
  \bibinfo{volume}{88}, \bibinfo{number}{2} (\bibinfo{date}{March}
  \bibinfo{year}{1998}), \bibinfo{pages}{315--326}.
\newblock
\showISSN{0377-0427}
\href{https://doi.org/10.1016/S0377-0427(97)00219-7}{doi:\nolinkurl{10.1016/S0377-0427(97)00219-7}}


\bibitem[Spiteri and Wei(2023)]%
        {spiteri2023fractional}
\bibfield{author}{\bibinfo{person}{Raymond~J. Spiteri} {and}
  \bibinfo{person}{Siqi Wei}.} \bibinfo{year}{2023}\natexlab{}.
\newblock \showarticletitle{Fractional-step {Runge--Kutta} methods:
  {R}epresentation and linear stability analysis}.
\newblock \bibinfo{journal}{\emph{J. Comput. Phys.}}  \bibinfo{volume}{476}
  (\bibinfo{year}{2023}), \bibinfo{pages}{111900}.
\newblock
\showISSN{0021-9991}
\href{https://doi.org/10.1016/j.jcp.2022.111900}{doi:\nolinkurl{10.1016/j.jcp.2022.111900}}


\bibitem[Strang(1968)]%
        {strang1968construction}
\bibfield{author}{\bibinfo{person}{Gilbert Strang}.}
  \bibinfo{year}{1968}\natexlab{}.
\newblock \showarticletitle{On the Construction and Comparison of Difference
  Schemes}.
\newblock \bibinfo{journal}{\emph{SIAM J. Numer. Anal.}} \bibinfo{volume}{5},
  \bibinfo{number}{3} (\bibinfo{date}{Sept.} \bibinfo{year}{1968}),
  \bibinfo{pages}{506–517}.
\newblock
\showISSN{1095-7170}
\href{https://doi.org/10.1137/0705041}{doi:\nolinkurl{10.1137/0705041}}


\bibitem[Suzuki(1992)]%
        {suzuki1992general}
\bibfield{author}{\bibinfo{person}{Masuo Suzuki}.}
  \bibinfo{year}{1992}\natexlab{}.
\newblock \showarticletitle{General Nonsymmetric Higher-Order Decomposition of
  Exponential Operators and Symplectic Integrators}.
\newblock \bibinfo{journal}{\emph{Journal of the Physical Society of Japan}}
  \bibinfo{volume}{61}, \bibinfo{number}{9} (\bibinfo{year}{1992}),
  \bibinfo{pages}{3015--3019}.
\newblock
\href{https://doi.org/10.1143/JPSJ.61.3015}{doi:\nolinkurl{10.1143/JPSJ.61.3015}}


\bibitem[Suzuki and Umeno(1993)]%
        {Suzuki:93}
\bibfield{author}{\bibinfo{person}{M. Suzuki} {and} \bibinfo{person}{K.
  Umeno}.} \bibinfo{year}{1993}\natexlab{}.
\newblock \showarticletitle{Higher-Order Decomposition Theory of Exponential
  Operators and Its Applications to {QMC} and Nonlinear Dynamics}. In
  \bibinfo{booktitle}{\emph{Computer Simulation Studies in Condensed-Matter
  Physics VI}}, \bibfield{editor}{\bibinfo{person}{David~P. Landau},
  \bibinfo{person}{K.~K. Mon}, {and} \bibinfo{person}{Heinz-Bernd
  Sch{\"u}ttler}} (Eds.). \bibinfo{publisher}{Springer Berlin Heidelberg},
  \bibinfo{address}{Berlin, Heidelberg}, \bibinfo{pages}{74--86}.
\newblock
\showISBNx{978-3-642-78448-4}
\href{https://doi.org/10.1007/978-3-642-78448-4_7}{doi:\nolinkurl{10.1007/978-3-642-78448-4_7}}


\bibitem[Tran et~al\mbox{.}(2024)]%
        {tran2024properties}
\bibfield{author}{\bibinfo{person}{Brian~K. Tran}, \bibinfo{person}{Ben~S.
  Southworth}, {and} \bibinfo{person}{Melvin Leok}.}
  \bibinfo{year}{2024}\natexlab{}.
\newblock \showarticletitle{On properties of adjoint systems for evolutionary
  pdes}.
\newblock \bibinfo{journal}{\emph{Journal of Nonlinear Science}}
  \bibinfo{volume}{34}, \bibinfo{number}{5} (\bibinfo{year}{2024}),
  \bibinfo{pages}{95}.
\newblock


\bibitem[Verner(1996)]%
        {verner1996high}
\bibfield{author}{\bibinfo{person}{JH Verner}.}
  \bibinfo{year}{1996}\natexlab{}.
\newblock \showarticletitle{High-order explicit Runge-Kutta pairs with low
  stage order}.
\newblock \bibinfo{journal}{\emph{Applied numerical mathematics}}
  \bibinfo{volume}{22}, \bibinfo{number}{1-3} (\bibinfo{year}{1996}),
  \bibinfo{pages}{345--357}.
\newblock


\bibitem[Walker and Ni(2011)]%
        {walker2011anderson}
\bibfield{author}{\bibinfo{person}{Homer~F. Walker} {and} \bibinfo{person}{Peng
  Ni}.} \bibinfo{year}{2011}\natexlab{}.
\newblock \showarticletitle{{A}nderson Acceleration for Fixed-Point
  Iterations}.
\newblock \bibinfo{journal}{\emph{SIAM J. Numer. Anal.}} \bibinfo{volume}{49},
  \bibinfo{number}{4} (\bibinfo{year}{2011}), \bibinfo{pages}{1715--1735}.
\newblock
\href{https://doi.org/10.1137/10078356X}{doi:\nolinkurl{10.1137/10078356X}}


\bibitem[Wan et~al\mbox{.}(2021)]%
        {wan2021quantifying}
\bibfield{author}{\bibinfo{person}{H. Wan}, \bibinfo{person}{S. Zhang},
  \bibinfo{person}{P.~J. Rasch}, \bibinfo{person}{V.~E. Larson},
  \bibinfo{person}{X. Zeng}, {and} \bibinfo{person}{H. Yan}.}
  \bibinfo{year}{2021}\natexlab{}.
\newblock \showarticletitle{Quantifying and attributing time step sensitivities
  in present-day climate simulations conducted with {EAMv1}}.
\newblock \bibinfo{journal}{\emph{Geoscientific Model Development}}
  \bibinfo{volume}{14}, \bibinfo{number}{4} (\bibinfo{year}{2021}),
  \bibinfo{pages}{1921--1948}.
\newblock
\href{https://doi.org/10.5194/gmd-14-1921-2021}{doi:\nolinkurl{10.5194/gmd-14-1921-2021}}


\bibitem[Yoshida(1990)]%
        {yoshida1990construction}
\bibfield{author}{\bibinfo{person}{Haruo Yoshida}.}
  \bibinfo{year}{1990}\natexlab{}.
\newblock \showarticletitle{Construction of higher order symplectic
  integrators}.
\newblock \bibinfo{journal}{\emph{Physics Letters A}} \bibinfo{volume}{150},
  \bibinfo{number}{5} (\bibinfo{year}{1990}), \bibinfo{pages}{262--268}.
\newblock
\showISSN{0375-9601}
\href{https://doi.org/10.1016/0375-9601(90)90092-3}{doi:\nolinkurl{10.1016/0375-9601(90)90092-3}}


\end{thebibliography}

% \appendix
% \section{Appendix Title}

\end{document}